# Advancements and Challenges in Quantum Machine Learning for Medical Image Classification: A Comprehensive Review

Md Farhan Shahriyar[a,] [*] and  Gazi Tanbhir[a]

[a]Department of Computer Science and Engineering , World University of Bangladesh,Dhaka,Bangladesh.

***Abstract –*** Quantum technologies are rapidly advancing as image classification tasks grow more complex due to large image volumes and extensive parameter updates required by traditional machine learning models. Quantum Machine Learning (QML) offers a promising solution for medical image classification. The parallelization of quantum computing can significantly improve speed and accuracy in disease detection and diagnosis. This paper provides an overview of recent studies on medical image classification through a structured taxonomy, highlighting key contributions, limitations and gaps in current research. It emphasizes moving from simulations to real quantum computers, addressing challenges like noisy qubits and suggests future research to enhance medical image classification using quantum technology..

*Keywords –Medical Image Classification, Quantum Machine Learning, Quantum Neural Network, Quantum Encoding*

## 1  INTRODUCTION

Since the development of quantum based computation as a transformative technology poised to significantly impact various sectors including healthcare. Within the domain of medical imaging, a critical area for diagnosis and treatment planning, conventional computing approaches often struggle with the complexity of data processing and analysis. Convolutional neural networks, one of the more traditional machine learning methods, are now proficient in finding patterns in medical imaging data. But the special powers of quantum systems have drawn interest because they may be able to overcome the constraints of traditional computing techniques [1]. Using concepts like superposition or entanglement in quantum computers are expected to find complex patterns that classical systems may struggle to handle efficiently potentially leading to breakthroughs in medical imaging tasks [2].

[*]Corresponding author
✉ farhanshahriyar.cse1@gmail.com (Md Farhan Shahriyar);

QML explores in which way quantum algorithms can surpass the capabilities of traditional machine learning approaches. This interdisciplinary field which combines quantum computing with machine learning is gaining interest in medical imaging due to its potential for tackling complex diagnostic challenges. Techniques like Quantum based Support Vector Machines, Quantum Convolutional Neural Networks also Variational Quantum Circuits leverage qubits—quantum information units that can process high-dimensional data faster than classical bits. In certain cases, quantum models such as Quantum K-Nearest Neighbors and Quantum Neural Networks outperform classical models by utilizing unique quantum features such as parallelism, entanglement, and superposition.

Additionally, techniques like Principal Component Analysis, commonly used in classical machine learning for dimensionality reduction, are being explored in quantum analogs such as Quantum Principal Component Analysis to better handle large medical datasets. Other approaches, like Quantum Annealing, have shown promise in optimization tasks related to medical imaging. However, quantum computing faces challenges in hardware reliability, error correction, and scalability. For instance, Parameterized Quantum Circuits and Quantum Tensor Networks offer flexible quantum architectures but require optimization to achieve accurate results.

The most current developments in applying quantum computing for medical imaging are reviewed in this study. It examines a number of QML techniques including variational quantum circuits, quantum support vector machines, and quantum convolutional neural networks. —and their use in medical imaging datasets including the Wisconsin Breast Cancer Database, COVIDx CXR3, and Medical MNIST. The review focuses on quantum encoding techniques, qubit architectures, and classification accuracies while addressing challenges like circuit parameterization and the scalability of quantum models for larger datasets.

A systematic search is conducted in prominent databases including IEEE Xplore, Google Scholar, and arXiv. The search utilized specific keywords such as "quantum," "image classification," and "medical image classification" to ensure a broad yet focused set of relevant studies. The search is limited to publications from the past two years up to 2024 to address recent advancements in the field.The goal of this review is to synthesize key findings on quantum machine learning's applications to medical image classification and to identify trends shaping future research. By highlighting both the potential and obstacles in the field, this QML study aims to provide a foundation for further exploration of quantum techniques like Quantum Variational Tensor Networks and other emerging methods for transforming medical imaging while addressing barriers to practical implementation.

## 2 OVERVIEW OF QUANTUM MACHINE LEARNING MEDICAL IMAGE CLASSIFICATION

The concept of a hybrid quantum-classical machine learning medical image classification approach is depicted in Fig 1.The classical components include data input, preprocessing, quantum state measurement, parameter optimization, and model output generation. Classical data is preprocessed before encoding into any quantum state. After quantum processing, measurements are carried out and classical optimization is implemented to update the quantum circuit parameters iteratively. The quantum part is to transform classical states of data into quantum states and process it by executing quantum circuits. The quantum circuit operates on qubits, exploring challenging state domains that classical systems find difficult to assess. The hybrid feedback loop pairs quantum and classical systems, allowing parameter updates via classical optimization following quantum processing for more effective model training.

A. Quantum Gates As it is easier to alter qubits, quantum gates are crucial to quantum computing. Superpositions are produced by the Hadamard gate (H), which converts $|0\rangle$ and $|1\rangle$ into $|0\rangle+|1\rangle$ and $|0\rangle-|1\rangle$, respectively. The Pauli X gate (X) can flips states and be compared as classical gate NOT . The Pauli-X Gate (X) is a form of classical NOT that reverses states. The Pauli-Y gate (Y) performs both flipping and phase shift. The Pauli-Z gate (Z) flips the phase of $|1\rangle$. The Toffoli gate (CCNOT) requires both control qubits to be set to $|1\rangle$ to flip the target qubit. The CNOT gate flips a target qubit based on the state of the control qubit. The gate is called SWAP can exchange two qubits at the same time of its states. Rotations around the Bloch sphere are performed using the rotation gates $Rx(\theta)$, $Ry(\theta)$, and $Rz(\theta)$. More specifically, the S and T gates provide $\pi 2$ phase shifts.

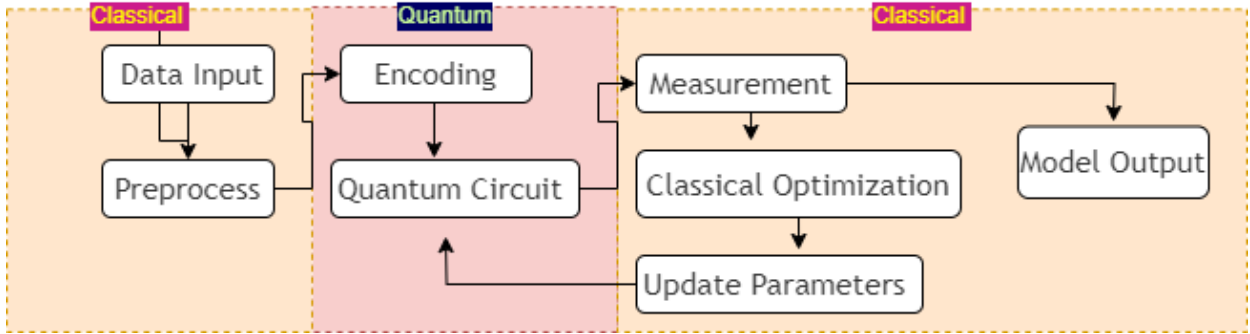

Fig. 1: A general quantum machine learning overview

B. Quantum Encoding Amplitude encoding maximizes circuit depth but serves the purpose of encoding large datasets by mapping amplitudes to qubits. The process of encoding classical data as qubit rotation angles is known as angle encoding. Dense Angle Encoding is the representation of the quantum state that reduces the number of qubits by applying rotations and phase shifts to combine two pixels into a single qubit. An angle encoding used in the study [3] has two RX gates (X-axis rotation) that are applied to each qubit. The rotation angles are connected to the input data that is being encoded into the quantum state. For angle encoding the quantum state is expressed by : $|\psi\rangle = \cos(\theta/2)|0\rangle + e^{i\phi}\sin(\theta/2)|1\rangle$. An input encoding circuit for malaria disease detection [3] is depicted in Fig. 3 which has a significant role in the detection accuracy. Four inputs, designated Q0 through Q3, are included, along with connections between the gates. Every input is routed via a Hadamard gate (H), then P gates with different exponents.

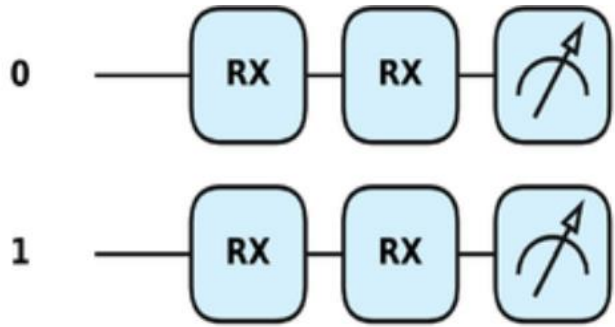

Fig. 2: Angle Encoding[3]

C. Variational Quantum Circuit Fig. 4 shows a quantum circuit with variational properties containing five qubits (q0 to q4). Every qubit rotates along the Y-axis and goes through an Ry gate, which has corresponding angles θ[0], θ[1], . . . , θ[4]. The parameters of the input data that are being encoded are reflected in these angles. In addition, the circuit has several controlled-phase gates, which are shown as dots joined by lines. As VQCs allow hybrid-classical training, the study [4] obtained 99.06% accuracy.

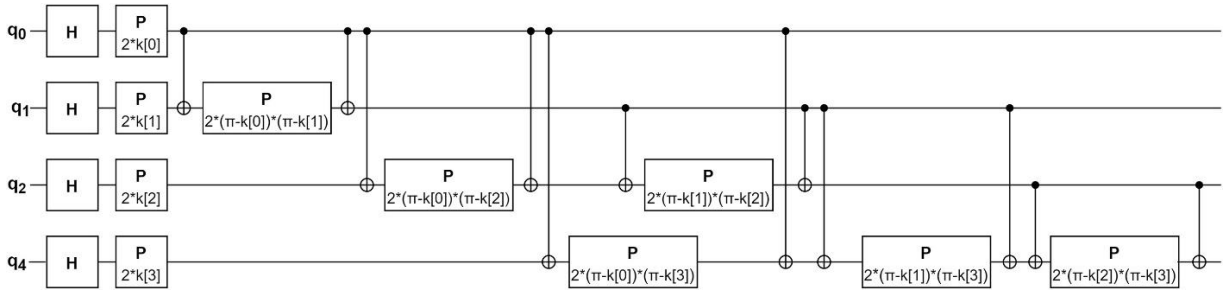

Fig. 3: Input Encoding Circuit [4]

The quantum circuit part shown in Fig. 5 can be replaced with a CNN architecture using convolution and pooling layers. Most research used ZFeatureMap to encode the classical data as spins around the Z-axis of qubits in quantum circuits. The study [3] presents remarkable empirical findings, indicating that rearrangment and different rotations can lead to significant differences in the results of the QCNN model. In this context, the encoded data is processed through the convolution and pooling layers by applying two unitary matrices with qubit parameters. Fig. 5 demonstrates the quantum convolution layer, which uses RX, RZ, and CNOT gates built from a series of parameterized two-qubit unitaries applied to neighboring qubit pairs. Fig. 6 describes quantum pooling layers that use RX, RZ, and CNOT gates to regulate entanglement. By parameterizing pooling, this entanglement can be simplified from a two-qubit circuit to a one-qubit unitary circuit using two random qubits. Half of the qubits are consolidated in a two-qubit pool at the quantum pooling layer, yielding the corresponding qubits labeled 1 for each state. The concept employs an image dataset for binary classification. When scaling an image to build features for input into a quantum network, individual pixel values are insufficient classification features.

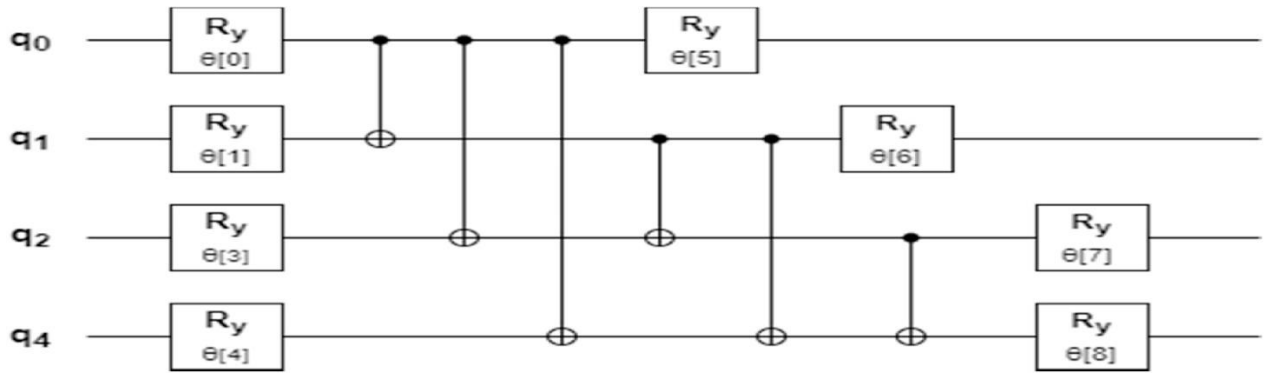

Fig. 4: The parameterized quantum circuit [4]

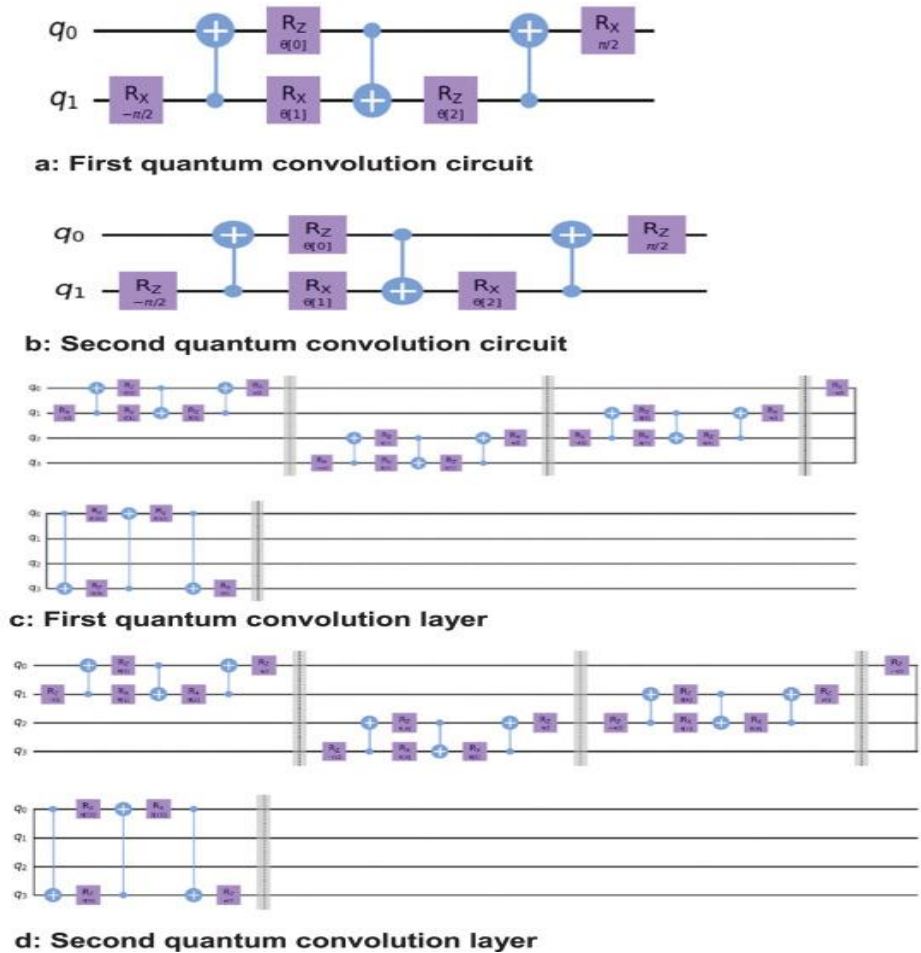

Fig. 5: Quantum convolution layers [3].

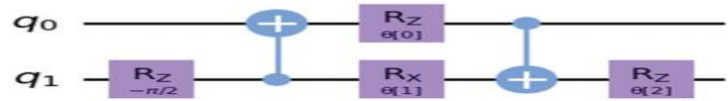

a: First quantum pooling circuit

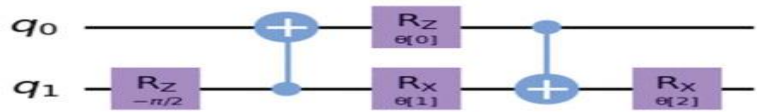

b: Second quantum pooling circuit

The
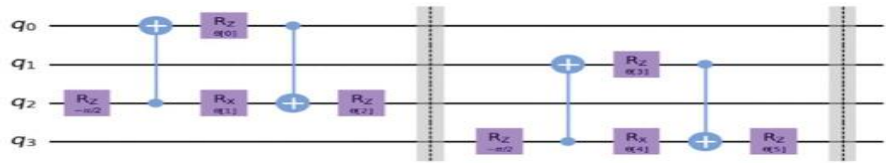

c: First quantum pooling layer

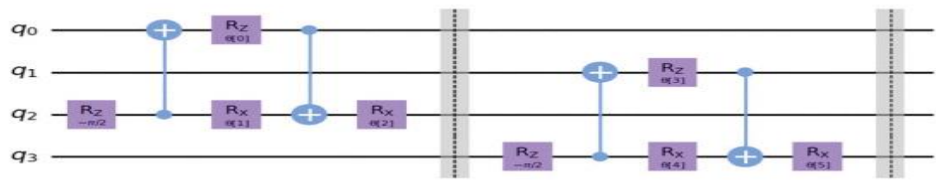

d: Second quantum pooling layer

Fig. 6: Quantum pooling layer [3].

Quantum Convolution Layer and Quantum Pooling Layer consist of two-qubit quantum circuits known

D. as QCNN1 and QCNN2, which have been customized for different quantum applications. Regardless of their similar architecture, the initial rotation gates used in QCNN1 and QCNN2 differ. The initial rotation gate selection (Rz vs. Rx) is likely to impact the evolution of the quantum state and entanglement patterns within the circuits. Consequently, one circuit design can surpass the other in terms of accuracy, gate count, or depth, depending on the individual quantum algorithm or task being performed.

# 3  STATE-OF-THE-ART QUANTUM MODELS

R V et al. [5] adopted procedures for extracting clinically significant characteristics from DBT images utilizing six different models and three dimensionality reduction strategies, resulting in crucial 2D and 8D features. The use of an angle embedding-based QKSVM classifier with a complex quantum kernel outperforms typical quadratic kernels in classifying DBT features. Fig. 6: Quantum pooling layer [3]. The LDA approach demonstrated that a smaller collection of 2D characteristics had more variance between classes. Their findings highlighted the significance of radiologist-defined areas of interest (ROIs) in improving classification accuracy and diagnostic clarity. The study demonstrated the use of quantum machine learning to improve DBT image categorization accuracy by building a quantum circuit with fewer qubits. Performance comparisons revealed that the QKSVM achieved 83.33% accuracy on a simulator, slightly higher than 82.8% on the AspenM3 quantum processing unit.

Hossain et al. [4] utilized a dataset of 27,558 Red Blood Cell images, which included equal numbers of malaria-infected and uninfected samples collected from microscopic blood smears. The study used a ZZ feature map for encoding, which transforms classical data into quantum states, and a 4-qubit variational quantum circuit for classification. Through 10-fold cross-validation, the proposed system achieved impressive performance metrics of 99.06% accuracy, 99.08% precision, 99.05% recall, and 99.07% specificity, surpassing traditional methods in both accuracy and feature extraction efficiency. The proposed quantum convolutional neural networks (QCNNs) for image classification [3] use quantum entanglement and efficient feature encoding to improve processing speed and accuracy. The architecture contains a quantum cluster state preparation layer that generates synthetic image data, as well as a ZFeatureMap that encodes classical data into quantum states via Z-axis rotations. The QCNN processes data using quantum convolution and pooling layers, with RX, RZ, and CNOT gates to handle entanglement. Two separate QCNN models were created, each with a different initial rotation gate, which influenced their performance.

The QTN-MLP model [6] is a medical image classification technique that combines tensor network methods with multilayer perceptron (MLP) architectures. At its foundation is the Matrix Product State (MPS), which successfully captures one-dimensional correlations within quantum states. QTN-MLP targets specific parts of a feature map, improving its capacity to extract important information while reducing interference from non-lesion areas. QTN-MLP creates a higherorder tensor that accurately reflects local features by segmenting the feature map and transforming it into onequbit quantum states. This tensor expands with the MPS to capture both spatial relationships and overall image characteristics. The design also includes Token MLP and Channel MLP components that aggregate information from various tokens and channels. Experimental results show that QTN-MLP outperforms traditional convolutional neural networks, achieving 89.2% accuracy on the COVID-CT dataset and 86.2% on the ISIC 2018 dataset while greatly minimizing model parameters. The research conducted [7] investigates Quantum Adaptive Machine Learning (QAML) for identifying diagnostic medical images with a specific emphasis on brain tumor MRI scans. QAML employs a hybrid quantum-classical neural network design that includes quantum convolutional and pooling layers followed by a classical fully connected layer. Experiments with two datasets: the Kaggle brain tumor dataset and the REMBRANDT dataset showed that QAML achieved convergence in only 63 epochs, surpassing classical models that needed 100 epochs to achieve comparable accuracy levels. The addition of a value-adjusted gradient descent optimizer increased convergence rates. The results revealed that QAML provides a considerable advantage in speed and efficiency, achieving up to 96.40% accuracy.

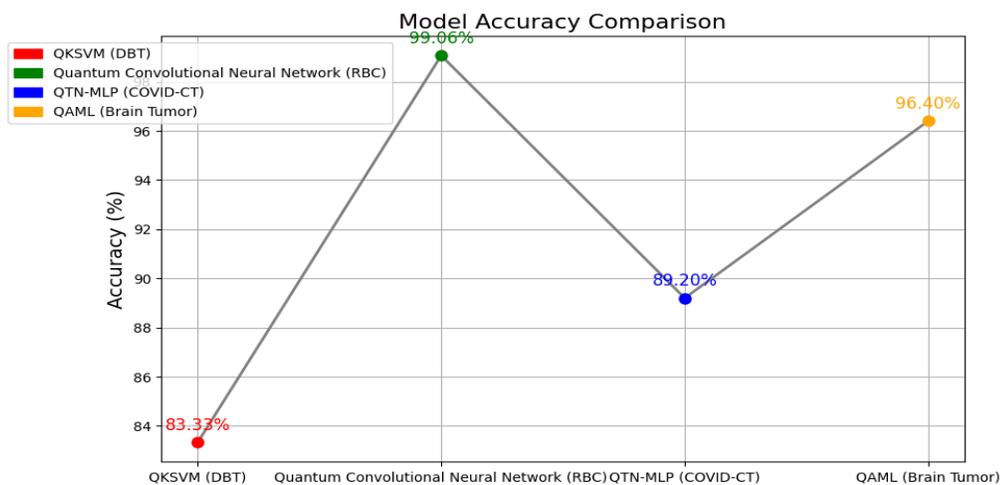

Fig. 7: State of the Art models performances

TABLE I: Summary of Quantum Machine Learning Applications in Medical Image Classification Papers

| Paper | Dataset | Encoding | Qubits | Result Accuracy | Limitation |
|---|---|---|---|---|---|
| Desai et al. [8] | UCI Machine Learning Repository's WBCD | Not Specified | Not Specified | QSVC 93%, VQC 78% | No specific information about encoding used, gates, and VQC architecture. |
| Yousif et al. [3] | COVIDx CXR-3 | ZFeatureMap (Angle) | 8-qubits | QCNN 94.21% | Ignored effectiveness of RZZ gates; did not present a stable, expressive model for capturing intricate data patterns. |
| Hassan et al. [9] | Medical MNIST | Threshold | 4-qubit | MQCNN 99.6% | Comparison with other works is missing; did not explore more encoding methods. |
| Reka et al. [10] | HAM10000 | Angle | 3-qubit | QSVC 72.5%, QNN 82.29% | "mel" and "nv" classes were poorly classified. |
| Khatoniar et al. [11] | Chest X-ray | Angle | 8-qubits | QCNN 73-75% | Lacked proper dataset description and evaluation of results. |
| Altares-Lopez et al. [12] | SARS-CoV-2, Brain Tumor | Angle | 5-qubits | PCA(2) + QSVM 96.7% [COVID], PCA(45) + QSVM 85.9% [Tumor] | QPCA could be used instead of PCA; results may vary on a real quantum device. |
| Mallick et al. [13] | BreakHis (Histopathological images) | Angle, Amplitude, Hybrid | Not Specified | SVC + VQC: RGB and HSV 100%, RGB and CIE L*u*v 95.4%; | Cross-validation not applied. |
| V et al. [?] | Digital Breast Tomosynthesis (DBT) | ZZFeatureMap (Angle) | Not Specified | QKSVM 84.48% | Larger dataset may disclose a broader study picture. |
| Ahmed et al. [?] | BreakHis (Histology images) | Angle | Not Specified | QOA 93.67% | Inadequate details about encoding method, QC architecture, and gates. |
| Suneel et al. [7] | Kaggle Brain MRI, REMBRANDT | Angle, Amplitude, Hybrid | Not Specified | QAML 69.46% - 96.40%; | Inadequate cross-validation raises concerns about overfitting and generalizability. |
| M Hossain et al. [4] | Kaggle Cell Image Malaria | ZZFeatureMap (Angle) | 4-qubit | PCA + mRMR + VQC 99.06% | Fewer parameters and limited feature set compared to other studies. |
| Odusami et al. [14] | ADNI (MRI and PET) | Not Specified | Not Specified | MViT: AD vs. CN 94.73%, AD vs. MCI 92.98%, CN vs. MCI 89.36% | Limited external validation datasets; reliance on decreased hippocampal volume can lead to misdiagnosis. |
| Simen et al. [15] | MedMNIST v2 (Breast Cancer and Pneumonia) | Angle | 4-qubit, 9-qubit | DAQCNN: 88.5% (4-qubit), 93.6% (9-qubit) | High execution time and untested scalability to larger diverse datasets. |
| Swathi et al. [16] | Algerian Ultrasound Images Thyroid Dataset (AUITD) | Angle | Not Specified | QuCNet: 95.74% (Tumor classification), 93.87% (suspicious-level) | Small dataset and reliance on hybrid quantum-classical computing lead to longer computation times. |
| Zhang & Lai [6] | COVID-CT, ISIC 2018 | Angle | 1-qubit | QTN-MLP: 89.2% (COVID-CT), 86.2% (ISIC 2018) | Imbalance in ISIC 2018 dataset; lack of testing on external datasets or cross-validation raises concerns about result credibility. |

# 4   MULTI-CLASS CLASSIFICATION IN QUANTUM MACHINE LEARNING FOR MEDICAL IMAGING

The field of quantum machine learning (QML) has traditionally concentrated on binary classification tasks. However, transitioning these methodologies to address multi-class classification represents a pivotal and demanding challenge. Multi-class classification in medical imaging involves distinguishing among three or more categories, each characterized by intricate and potentially overlapping features. This task is further complicated by several factors, including the inherent imbalance in medical datasets, the high computational requirements of processing numerous classes, and the complex relationships that often exist in high-dimensional feature spaces. As QML evolves, addressing these issues is critical for its application to diverse and realistic clinical scenarios.

A. Leveraging Quantum Neural Architectures for Multiclass Classification

Quantum neural networks, such as Quantum Convolutional Neural Networks (QCNNs), have shown substantial promise in multi-class classification due to their ability to process high-dimensional data efficiently. By leveraging quantum phenomena like entanglement and superposition, QCNNs can extract hierarchical features from medical imaging data, capturing intricate patterns that might elude classical convolutional networks. For instance, recent studies have employed QCNNs to classify histopathological images into multiple disease categories, achieving higher accuracy than classical convolutional models in specific scenarios [11]. The primary advantage of QCNNs lies in their potential to represent and process complex data distributions through quantum states, which may reduce the need for extensive classical preprocessing. However, the practical implementation of QCNNs faces challenges, such as managing quantum noise, ensuring circuit scalability, and addressing the limitations imposed by current quantum hardware. These issues necessitate the optimization of quantum circuit design and the development of noise resilient algorithms to maximize their applicability in multi-class classification tasks.

B. Parameterized Quantum Models for Complex Decision Boundaries

Parameterized quantum models, particularly Variational Quantum Circuits (VQCs), are increasingly being explored for their ability to handle multi-class classification challenges. VQCs utilize parameterized quantum gates that can adapt to represent complex decision boundaries within datasets. This adaptability makes them well-suited for multi-class problems in medical imaging, where features often exhibit non-linear separability. One noteworthy application involved the classification of multiple cancer subtypes in magnetic resonance imaging (MRI) datasets. VQCs demonstrated competitive accuracy with fewer trainable parameters compared to their classical counterparts, highlighting their efficiency in parameter optimization [17]. This reduced parameter requirement also translates into lower memory and computational demands, which is a critical consideration when dealing with large medical datasets. Despite their promise, VQCs face notable limitations, such as the difficulties associated with variational algorithm optimization, including barren plateaus and the sensitivity of quantum hardware to noise. Moreover, the optimization process can become increasingly complex as the number of classes grows, emphasizing the need for tailored techniques to balance performance and computational feasibility.

C. Hybrid Quantum-Classical Frameworks for Enhanced Accuracy

Hybrid quantum-classical frameworks, such as Quantum Transfer Learning (QTL), represent an innovative approach to overcoming the limitations of standalone quantum or classical models. These frameworks combine the strengths of quantum computing with the robustness of pre-trained classical models, enabling the effective transfer of learned representations. In the domain of medical imaging, QTL has been utilized to classify diabetic retinopathy into multiple stages, using quantum computational layers to refine decision making processes while maintaining the interpretability and robustness of classical methods [18]. The integration of quantum layers within classical architectures offers several advantages. First, it allows for the extraction of quantum-enhanced features that are not readily accessible through classical methods. Second, hybrid models can efficiently scale to multi-class problems by leveraging the computational efficiency of classical systems for initial feature extraction and utilizing quantum circuits for complex pattern recognition. However, these approaches introduce additional complexities, such as ensuring seamless integration between quantum and classical components, optimizing hybrid parameters and addressing potential bottlenecks in the interaction between the two paradigms.

# 5 REAL-WORLD APPLICATIONS USING QUANTUM HARDWARE

While much of the research in QML has been conducted using simulations, recent advances have enabled the application of quantum algorithms to real-world hardware. For example, IBM's quantum processors have been employed to perform breast cancer classification demonstrating the viability of small-scale medical imaging tasks on current quantum systems. Similarly, Google's Sycamore processor has been utilized for skin lesion classification achieving encouraging results despite the constraints of contemporary quantum devices. The application of QML to actual quantum hardware underscores several challenges, including quantum noise, a limited number of available qubits, and high error rates inherent in noisy intermediate-scale quantum (NISQ) devices. To address these issues, researchers have explored techniques like quantum error correction and error mitigation. For instance, zero-noise extrapolation has been used to enhance the reliability of quantum classifiers, allowing for more accurate predictions in noisy environments [19].

Despite these challenges, real-world applications of quantum hardware continue to advance highlighting the growing potential of QML in clinical contexts. In comparison to traditional machine learning methods as the advancement grows, it increases processing speed, reduces model complexity and improves image classification precision. Future research should focus on hardware specific optimizations and the development of robust algorithms capable of mitigating noise and scaling effectively as quantum devices continue to mature.

# 6 COMPARATIVE ANALYSIS OF QUANTUM AND CLASSICAL METHODS

A comparative analysis of quantum and classical methods in medical imaging reveals distinct strengths and limitations for each approach. Quantum methods excel in handling high-dimensional data and complex feature spaces, offering potential advantages in applications such as image segmentation, anomaly detection, and multi-class classification. Classical methods, on the other hand, benefit from mature ecosystems, extensive optimization, and accessibility to researchers without specialized quantum expertise. While quantum algorithms hold the promise of exponential or polynomial speedups for certain tasks, their practical application is often hindered by hardware constraints, such as limited qubit availability and high error rates. Conversely, classical methods provide reliable and scalable solutions for large datasets but may struggle with computational bottlenecks when dealing with complex or high-dimensional datasets. The integration of both paradigms through hybrid models represents a promising avenue to take advantage of their respective strengths [20].

# 7  DISCUSSION AND FINDINGS

Quantum models, including QCNNs, VCNNs, and QSVMs, provide immense promise in outperforming traditional approaches for specific tasks. However, significant problems remain, particularly in widening the models' applicability to more complicated tasks, improving hardware capabilities, and perfecting image encoding methods.

Future studies may highlight the following:

• RGB Image Support: Current quantum models have major difficulties when processing RGB images, which is critical for broadening their applicability in other domains. Future research should concentrate on improving the ability of quantum models to handle multi-channel (color) images successfully.

• Multi-Class Classification: While several quantum models have proven successful in binary classification tasks, their performance in complicated multiclass classification contexts remains unexplored. Future studies must develop quantum models to handle increasingly complex classification problems, notably in the context of VQC, QAML, and QTN-MLP implementations.

• Noise-Resistant Circuits: Given the inherent noise sensitivity of current quantum hardware, the development of noise-resistant quantum circuits is vital for practical use in the real world. Addressing this difficulty will be critical to improving the validity and precision of quantum models.

• Efficient Image Encoding: Scaling quantum models for larger images is a significant barrier for efficient encoding. Downscaling frequently leads to the loss of critical information, emphasizing the need for novel encoding approaches that preserve image integrity while remaining compatible with quantum processing.

# 8　CONCLUSION

This study explores the emerging field of medical image classification, where there is a notable shift from classical machine learning methods to quantum machine learning (QML). Recent research has predominantly focused on Quantum Convolutional Neural Networks (QCNNs) and Quantum Support Vector Machines (QSVMs), which have shown promise in addressing classification tasks in medical imaging. This review provides a comprehensive overview of the current literature, highlighting key aspects such as the number of qubits used, quantum encoding techniques, and the inherent limitations of existing models. It also identifies areas underexplored within medical image classification that warrant further investigation, including the potential for more diverse datasets, improved optimization methods, and alternative quantum models. By pinpointing these research gaps, this study aims to stimulate future advancements in quantum machine learning for medical imaging, paving the way for more efficient and accurate diagnostic tools in clinical settings. As the field progresses, addressing these challenges will be essential to fully realize the potential of QML in practical, real-world applications.

.